\documentclass{article} 
\usepackage{mhchem}
\usepackage{makecell}
\usepackage{iclr2026_conference,times}

\usepackage{amsmath,amsfonts,bm}

\def\eqref#1{equation~\ref{#1}}

\def\1{\bm{1}}

\def\ra{{\textnormal{a}}}

\def\rx{{\textnormal{x}}}

\def\rva{{\mathbf{a}}}

\def\erva{{\textnormal{a}}}

\def\ervx{{\textnormal{x}}}

\def\rmA{{\mathbf{A}}}

\def\vmu{{\bm{\mu}}}
\def\vtheta{{\bm{\theta}}}
\def\va{{\bm{a}}}

\def\ve{{\bm{e}}}

\def\vx{{\bm{x}}}

\def\eva{{a}}

\def\mA{{\bm{A}}}

\def\mH{{\bm{H}}}
\def\mI{{\bm{I}}}
\def\mJ{{\bm{J}}}

\def\mX{{\bm{X}}}

\def\mSigma{{\bm{\Sigma}}}

\DeclareMathAlphabet{\mathsfit}{\encodingdefault}{\sfdefault}{m}{sl}
\SetMathAlphabet{\mathsfit}{bold}{\encodingdefault}{\sfdefault}{bx}{n}
\newcommand{\tens}[1]{\bm{\mathsfit{#1}}}
\def\tA{{\tens{A}}}

\def\tX{{\tens{X}}}

\def\gG{{\mathcal{G}}}

\def\sA{{\mathbb{A}}}
\def\sB{{\mathbb{B}}}

\def\sS{{\mathbb{S}}}

\def\emA{{A}}

\newcommand{\etens}[1]{\mathsfit{#1}}

\def\etA{{\etens{A}}}

\newcommand{\E}{\mathbb{E}}

\newcommand{\R}{\mathbb{R}}

\newcommand{\KL}{D_{\mathrm{KL}}}
\newcommand{\Var}{\mathrm{Var}}

\newcommand{\Cov}{\mathrm{Cov}}

\newcommand{\normltwo}{L^2}
\newcommand{\normlp}{L^p}

\newcommand{\parents}{Pa} 

\usepackage{hyperref}
\usepackage{url}
\usepackage{comment}
\usepackage{graphicx}
\usepackage{subcaption}
\usepackage{booktabs}
\usepackage{adjustbox}
\usepackage{textcomp}

\title{A Comparative Study of Molecular Dynamics Approaches for Simulating Ionic Conductivity in Solid Lithium Electrolytes}

\author{%
Dounia Shaaban Kabakibo \\
D\'epartement de physique, Universit\'e de Montr\'eal \\
Mila -- Quebec AI Institute \\
Montr\'eal, Canada \\
\texttt{dounia.shaaban.kabakibo@umontreal.ca}
\And
F\'elix Therrien \\
Mila -- Quebec AI Institute \\
Montr\'eal, Canada
\And
Yoshua Bengio \\
D\'epartement d'informatique et de recherche op\'erationnelle, Universit\'e de Montr\'eal \\
Mila -- Quebec AI Institute \\
Montr\'eal, Canada
\And
Michel C\^ot\'e \\
D\'epartement de physique, Universit\'e de Montr\'eal \\
Mila -- Quebec AI Institute \\
Montr\'eal, Canada
\And
Hongyu Guo \& Homin Shin \\
National Research Council Canada \\
Ottawa, Canada
\And
Alex Hernandez-Garcia \\
D\'epartement d'informatique et de recherche op\'erationnelle, Universit\'e de Montr\'eal \\
Mila -- Quebec AI Institute \\
Institut Courtois, Universit\'e de Montr\'eal \\
Montr\'eal, Canada
}

\iclrfinalcopy 
\begin{document}

\maketitle

\begin{abstract}
Accurate prediction of ionic conductivity is critical for the design of high-performance solid-state electrolytes in next-generation batteries. We benchmark molecular dynamics (MD) approaches for computing ionic conductivity in 21 lithium solid electrolytes for which experimental ionic conductivity has been previously reported in the literature. In particular, we compare simulations driven by density functional theory (DFT) and by universal machine-learning interatomic potentials (uMLIPs), namely a MACE foundation model. We find comparable performance between DFT and MACE, despite MACE on one GPU more than 350 times faster than DFT on a 64-CPU node. The framework developed here is designed to enable systematic comparisons with additional uMLIPs and fine-tuned models in future work.
\end{abstract}

\section{Introduction}
\label{sec:introduction}
Solid-state electrolytes are a key component of next-generation energy storage technologies, offering the promise of improved safety and higher energy density compared to liquid electrolytes \citep{kim2015review}. A central property governing their performance is ionic conductivity, which quantifies how efficiently mobile ions migrate through a solid under an applied electric field \citep{mehrer2007diffusion}.
While experimental measurements remain the reference standard, computational approaches play an important role in screening candidate materials and providing physics-based insights into ion transport.
Molecular dynamics (MD) simulations offer a direct route to computing ionic conductivity from atomic trajectories, but the reliability of the results depends critically on the quality of the force model and the accessible system sizes and simulation times. 

Ab initio molecular dynamics based on density functional theory (DFT) provide a principled description of interatomic forces, yet its high computational cost severely limits the length and scale of simulations. Recent advances in universal machine-learning interatomic potentials (uMLIPs) have opened new possibilities for large-scale MD simulations. Models such as the MACE foundation model \citep{batatia2023foundation} are trained on diverse DFT data and can, in principle, be applied across a wide range of chemistries, on systems involving thousands of atoms at the scale of nanoseconds and in a fraction of the time needed for DFT simulations. However, it remains unclear how accurately MD simulations driven by uMLIPs reproduce emergent transport properties such as ionic conductivity. This motivates a systematic benchmarking against ab initio molecular dynamics.

\subsection{Contributions}
Our approach introduces \textbf{three key contributions}:

\begin{enumerate}
    \item \textbf{A consistent dataset:} We consider a relatively large set of materials for which ionic conductivities have been experimentally measured, enabling a direct and systematic comparison between simulations and experiments.
    
    \item \textbf{A unified procedure:} We propose a unified procedure for ionic conductivity estimation that standardizes parameter choices and uncertainty quantification, providing a statistically meaningful basis for data-driven comparison across methods.

     \item \textbf{A comparison of multiple simulation approaches:} With this dataset and our proposed procedure, we assess the performance of two simulation methods—ab initio molecular dynamics as well as a general uMLIP—and how they compare to experimental measurements using an identical simulation protocol.
\end{enumerate}
Previous work with connections to ours is reviewed in Appendix~\ref{app:related_work}.

\section{Methods}
\label{sec:methods}
\subsection{Ionic Conductivity from Molecular Dynamics Simulations} 
\label{sec:ionicconductivitymd}
Room temperature ionic conductivity $\sigma$ is a key property that characterizes ion transport in solid state materials. One approach to estimate $\sigma$ is to use MD simulations, which track the time evolution of atoms in a material based on Newton’s equations of motion. These simulations allow the study of ionic diffusion at finite temperatures by modeling atomic interactions through different levels of theory: DFT, empirical potentials, or MLIPs. After performing an MD simulation, the self diffusion coefficient $D$ for mobile ions (in this case, lithium ions) is obtained using the Einstein relation
\begin{equation} \label{eq:msd}
D = \lim_{t \to \infty} \frac{\text{MSD}(t)}{2d\,t}, 
\end{equation}
where $d=3$ is the system's dimensionality and $\text{MSD}( t)$ is the per ion mean square displacement at time $ t$. In practice, $D$ is calculated from the slope of the $\text{MSD}(t)$ plot over a finite time window. 
Once $D$ is determined, the ionic conductivity $\sigma$ is obtained via the Nernst-Einstein relation
\begin{equation}
\label{eq:ic}
\sigma (T) = \frac{N q^2 D(T)}{V k_B T},
\end{equation}
where $N$ is the total number of lithium ions, $q$ is the charge of the diffusing ion, $V$ is the cell volume, $k_B$ is the Boltzmann constant and $T$ is the temperature.
Because room-temperature ionic conductivity requires long MD timescales, which are computationally prohibitive, especially for ab initio MD (AIMD), simulations are instead performed at higher temperatures. The diffusion coefficients estimated at these temperatures are used to fit the Arrhenius equation:
\begin{equation}
\label{eq:arrhenius}
D(T)=D_0\exp{\left (-\frac{E_a}{k_BT}\right )},
\end{equation}
where $E_a$ is the activation energy. This relation is then extrapolated to estimate the conductivity at room temperature. Despite the widespread use of the Arrhenius fitting, strict Arrhenius behavior is not universal and deviations have been observed \citep{qi2021bridging}.


\subsection{Materials and parameters choice}
A set of 21 solid lithium electrolytes covering a wide range of ionic conductivities were selected from the OBELiX database \citep{therrien2025obelix} which is a collection of synthesized solid electrolytes and their experimentally measured room-temperature conductivities compiled from the literature. Further details are provided in Appendix~\ref{app:materials}. 

 We considered two classes of force calculators: fully \textit{ab initio} (DFT) and uMLIPs. In the present work, we focus on a single uMLIP, namely the MACE model \citep{batatia2023foundation}, using the \texttt{medium-mpa-0} checkpoint.

In our procedure, MD simulations were performed on supercells with minimum dimensions of $10 \mathrm{\AA} \times 10 \mathrm{\AA} \times 10 \mathrm{\AA}$ \footnote{Exceptions were made for AIMD simulations. See Appendix~\ref{app:aimd+ft}.}. From these simulations, for each material, we obtained the diffusion coefficient $D$ at 5 different temperatures that we then used to fit \eqref{eq:arrhenius} and obtain the extrapolated room temperature diffusion coefficient $D(T=300\text{K})$. Simulations were carried out at temperatures ranging from 800 to 1200~K in 100~K increments in the NVT ensemble using a Nosé--Hoover thermostat and a time step of 2~fs. Each MD trajectory was 100~ps long, with the first 5~ps discarded prior to diffusion analysis. MD simulations using MACE models were carried out with the Atomic Simulation Environment (ASE) \citep{larsen2017atomic}. Parameters used for AIMD simulations are detailed in Appendix~\ref{app:aimd+ft}.

\begin{figure}[h]
\centering
\includegraphics[width=\linewidth]{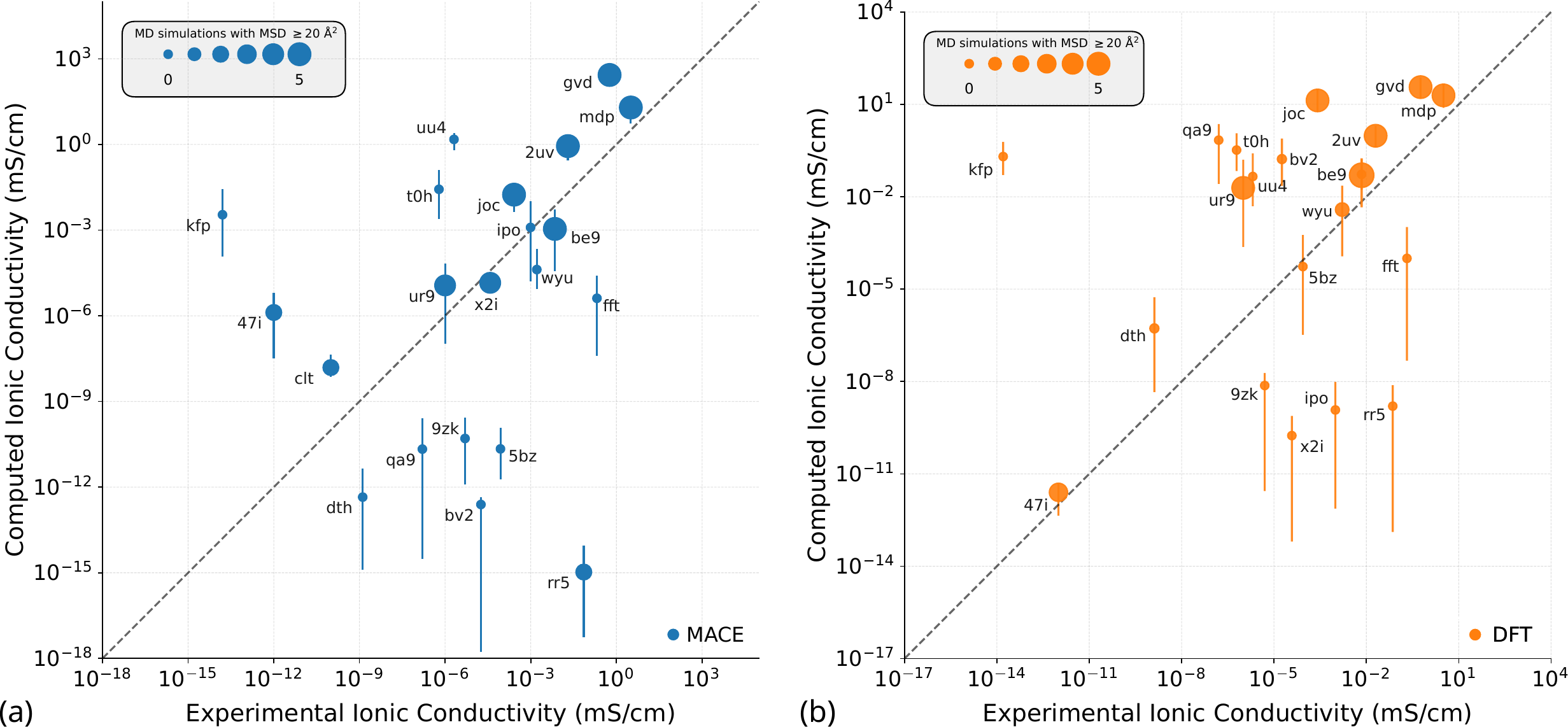}
\caption{Parity plot comparing ionic conductivity computed using (a) MACE and (b) DFT foundation model with experimental values reported by \citet{therrien2025obelix}. The size of the circles reflect the number of simulations (at 5 different temperatures) that reached at least $\mathrm{MSD} = 20~\mathrm{\AA}^2$.} 
\label{fig:dft_mace_vs_exp}
\end{figure}

Uncertainty estimation for diffusion coefficients was carried out using the \texttt{Kinisi-2.0.1} package \citep{mccluskey2024accurate, mccluskey2024kinisi}. In this framework, ionic diffusion is statistically analyzed by modeling atomic displacements over time and propagating their uncertainties to obtain confidence intervals on the diffusion coefficient. 
The parameters used  in the documentation examples were used throughout and no ablation study was conducted to assess sensitivity to these settings. 

For each material, as a measure of reliability, we also report how many of the five simulations (one for each temperature) had at least some ionic movement ($\mathrm{MSD} = 20~\mathrm{\AA}^2$). This threshold corresponds to a characteristic displacement of $\sim 4.5~\mathrm{\AA}$, a length scale comparable to typical interatomic distances in inorganic crystals, indicating motion beyond local vibrational fluctuations.




\section{Results}
 In this section, we present results for the analysis discussed above. Figure~\ref{fig:dft_mace_vs_exp} shows the parity plot of the estimated ionic conductivity using both DFT\footnote{DFT simulations for 1 material could not be completed within a reasonable amount of time.} and the MACE foundation model versus the experimental values included in \citet{therrien2025obelix}. The materials are labeled with their corresponding names in OBELiX. The chemical formulas are given in Appendix~\ref{app:materials}. 

The figures also show how many of the MD simulations at 5 temperatures had $\mathrm{MSD}$ above threshold for each material. We observe that most materials passing all checks have estimated conductivities close to the experimental values. Conversely, the estimations from simulations that fail to meet the displacement threshold tend to diverge from the experimental values. Interestingly, some materials which are known to exhibit high diffusion experimentally, namely \texttt{fft} ($\beta$-\ce{Li3N}) and \texttt{rr5} (\ce{LiTi2(PO4)3}), show little diffusion in our simulations. We anticipate that a diffusion mechanism mediated by a small concentration of defects---either not significant enough to be reported in lab experiments or not directly measurable---is necessary for diffusion in these cases. Future work will extend this analysis. 
Additionally, it should be noted that the experimental value $\sigma = 10^{-12}$~mS/cm we use for \texttt{47i} (Li$_4$P$_2$O$_7$) may be lower than the actual value, as the literature only reports an upper bound ($\sigma_{\text{exp}} < 10^{-7}$~mS/cm).
Figures providing additional insight on the previous results can be found in Appendix~\ref{app:mace_vs_dft+err}. 

In Table~\ref{tab:resources}, we report details about the computational resources used by both methods. For the DFT simulations, the number of CPUs used for each material was optimized based on system size (see Appendix~\ref{app:aimd+ft}). For the MACE simulations, a single GPU was used for all materials using a \texttt{1x NVIDIA A100-SXM4-40GB} paired with one CPU core. Each reported value corresponds to the computational cost of a single MD simulation (one material at one temperature). Since MD simulations at different temperatures are independent, they can be run concurrently provided sufficient computational resources are available. The reported per-simulation runtimes therefore approximate the expected wall-clock time of one  MD simulation under parallel execution and optimized resources.

\begin{table}[t]
\caption{Aggregated computational cost over all materials per temperature. Times are reported in \texttt{dd:hh:mm}. Only maximum available GPU memory usage is reported.}
\label{tab:resources}
\begin{center}
\setlength{\tabcolsep}{4pt}
\begin{tabular}{c @{\hspace{8pt}} cc @{\hspace{8pt}} ccc @{\hspace{8pt}} cc}
\toprule
\textbf{Method} 
& \multicolumn{2}{c}{\textbf{\# of devices}} 
& \multicolumn{3}{c}{\textbf{Total time}} 
& \multicolumn{2}{c}{\textbf{Max. mem. (GB)}} \\ 
& GPU & CPU 
& Median & Min & Max
& GPU & CPU \\
\midrule
MACE 
& 1 
& 1 
& 47m & 20m & 2h 27m 
& 40
& 1.2 \\
DFT
& 0 
& 64 
& 9d 21h 09m & 2d 3h 0m & 42d 09h27m
& 0
& 32  \\
\bottomrule
\end{tabular}
\end{center}
\end{table}
\section{Discussion and conclusion}
While the number of materials considered in this study remain limited to draw robust statistical conclusions, several trends emerge. First, molecular dynamics simulations using MACE require only a fraction of the computational time needed for DFT-based simulations, provided that GPU resources are available. Specifically, MACE on one GPU is on average 378 times faster than DFT on one 64-CPU node. Although DFT simulations can in principle also be accelerated on GPUs, the expected speedup is not as substantial as for uMLIPs. Moreover, many researchers in the materials science community still rely primarily on CPU-based DFT calculations, which further motivates our DFT simulations being CPU--based. 

Despite this large difference in computational cost, MACE and DFT exhibit comparable predictive performance. Quantitatively, we evaluate performance by computing the mean absolute error (MAE) between the logarithm of the predicted and the experimentally measured ionic conductivities. Over the 21 materials studied with MACE, we obtain an MAE of 4.11. For the 20 materials for which DFT results are available, the MAE is 4.10. Restricting the analysis to the materials passing all five MSD reliability checks --- five for MACE and six for DFT --- the MAE decreases to 1.55 for MACE and 2.35 for DFT. It is important to emphasize that molecular dynamics results become severely unreliable when little to no ionic motion is observed, as the extracted transport properties in such cases mainly reflect thermal fluctuations rather than genuine diffusion. Although the statistics over the five most reliable materials remain limited, these results suggest that MACE achieves performance comparable to DFT. 

To put these results in perspective, \citet{therrien2025obelix} found that the mean absolute variation over experimental ionic conductivity measurements of the same material was about 0.4 orders of magnitude and that a random forest model trained on the OBELiX dataset could achieve an MAE of about 1.6 on their test set.

These findings do not diminish the value of molecular dynamics simulations, which remain essential for understanding diffusion mechanisms and the underlying physics of ionic transport. However, they suggest that, in a high-throughput screening context, MD simulations should be performed selectively: primarily using uMLIPs such as MACE, and only occasionally with DFT when higher-fidelity calculations are strictly necessary.
Beyond screening, these results are also  particularly relevant in the context of machine-learning–driven materials discovery. In practice, the amount of high-quality experimental data on solid-state electrolytes remains limited, and models trained to predict ionic conductivity directly might exhibit restricted transferability to unseen chemistries. In this setting, multi-fidelity active learning frameworks \citep{hernandez2023multi} that combine large numbers of low-cost uMLIP-based MD simulations with a smaller number of higher-fidelity DFT calculations, offer a promising path forward. Future work directions are discussed in Appendix~\ref{app:future_work}.

\section*{Acknowledgments}
The authors thank Gabor Csanyi and Mickael Dolle for valuable discussions. The authors acknowledge support from the National Research Council Canada (NRC) through a collaborative R\&D grant (AI4D-core-132), Calcul Québec and the Digital Research Alliance of Canada. D.S.K. acknowledges support by the NSERC Postgraduate Scholarships – Doctoral program and the FRQ Nature and technologies Sector – Doctoral Research Scholarship Program. M.C. is a member of the Regroupement québécois sur les matériaux de pointe (RQMP).
\newpage

\bibliography{iclr2026_conference}
\bibliographystyle{iclr2026_conference}

\appendix
\section{Appendix}
\subsection{Related work}
\label{app:related_work}
High-throughput and machine-learning-based approaches are now central to the efforts for the discovery of solid ionic conductors. Early screening frameworks relied on workflows based on physics and molecular dynamics to estimate lithium diffusion across large materials databases \citep{kahle2020high}. Machine learning was subsequently introduced as a data-driven pre-screening tool to prioritize promising Li-ion conductors and reduce the cost of first-principles simulations \citep{sendek2018machine}, and later combined with cloud high-performance computing to enable screening of tens of millions of candidates with experimental validation of a few selected solid electrolytes \citep{chen2024accelerating}. More recently, machine-learned interatomic potentials have been used to model ion diffusion directly through MD in representative electrolytes \citep{fragapane2025li}, while other studies have leveraged uMLIPs or ML models to infer transport trends \citep{maevskiy2024predicting} or predict migration barriers \citep{dembitskiy2025benchmarking}, without running MD. Generative modeling has also been used to accelerate MD simulations in Lithium solid electrolytes \citep{nam2025flow}.

\subsection{Future work}
\label{app:future_work}
While the present results focus on simulations using a MACE uMLIP, the full framework---intended to be explored in future work---is designed to enable consistent comparisons across additional approaches, including other uMLIPs as well as material-specific models fine-tuned on DFT data. Such study would provide a systematic way to quantify the performance of uMLIPs and their fine-tuning strategies for long-time-scale diffusion simulations involving large numbers of atoms, and how they compare to fully ab initio simulations.
The present study is limited to fully occupied crystal structures. In many solid electrolytes, however, ionic diffusion is known to be mediated by defects. Extending this work to partially occupied structures, such as those available in the OBELiX database, would therefore be of significant interest. In addition, one could consider relaxing the experimental structures prior to running MD simulations to assess the sensitivity of the computed conductivities to structural relaxation. Additional factors not examined here include the impact of longer simulation times and the number of temperature points used in the Arrhenius fits. Finally, introducing criteria to detect possible structural melting would be an important consideration, especially for simulations conducted at high temperatures.

\subsection{Materials Selection}
\label{app:materials}
As mentioned before, the materials were selected from the OBELiX database \citep{therrien2025obelix}. As a first filter, only materials with fully occupied sites were considered to avoid complexities related to partial occupancies. For this subset, the range of ionic conductivities was divided on a logarithmic scale into 20 equal bins, from $\sigma = 10^{-7}$~mS/cm$^2$ to the highest measured value in the dataset, 25~mS/cm$^2$. Two bins were empty, and one material was randomly selected from each of the remaining 18 bins. To include poorly conducting materials, the lower conductivity range, $\sigma = 10^{-15}$–$10^{-7}$~mS/cm$^2$, was further divided into four equal bins, from which one material per bin was selected, resulting in a total of 22 materials. MD simulations using the MACE uMLIP failed for \ce{LiClC3H7NO}, and therefore this material was excluded from further analysis. This procedure resulted in a total of 21 materials. Table~\ref{tab:selected_materials} gives the chemical formulas of the selected materials.
Cross-referencing our material selection against the MPtraj \citep{deng_2023_chgnet} and sAlexandria \citep{barroso2024open} datasets on which the Mace \texttt{medium-mpa-0} model was trained returned matches for 11 Obelix IDs within the 5\% lattice parameter and 5\textdegree{} angle difference thresholds: \texttt{qa9}, \texttt{bv2}, \texttt{be9}, \texttt{5bz}, \texttt{gvd}, \texttt{fft}, \texttt{9zk}, \texttt{47i}, \texttt{clt} , and \texttt{joc}.
\begin{table}[t]
\caption{Selected materials and their chemical formulas.}
\label{tab:selected_materials}
\begin{center}
\begin{tabular}{lll}
OBELiX ID & Reduced chemical formula  \\
\hline
qa9 & $\alpha$-Li$_3$BN$_2$            \\
t0h & Li$_2$ZrO$_3$            \\
ur9 & LiZr$_2$(PO$_4$)$_3$      \\
uu4 & LiVO$_3$                 \\
9zk & Li$_5$GaO$_4$             \\
bv2 & LiYO$_2$                 \\
x2i & LiBiO$_2$               \\
5bz & $\beta$-Li$_3$BN$_2$              \\
joc & Li$_3$PS$_4$              \\
ipo & Li$_7$La$_3$Hf$_2$O$_{12}$  \\
wyu & Li$_7$La$_3$Zr$_2$O$_{12}$  \\
be9 & LiGaBr$_4$               \\
2uv & Li$_4$SnSe$_4$            \\
rr5 & LiTi$_2$(PO$_4$)$_3$      \\
goi & LiClC$_3$H$_7$NO          \\
fft & $\beta$-Li$_3$N                  \\
gvd & $\alpha$-Li$_3$N                  \\
mdp & Li$_7$P$_3$S$_{11}$       \\
dth & Li$_4$GeO$_4$             \\
clt & Li$_6$CuB$_4$O$_{10}$     \\
47i & Li$_4$P$_2$O$_7$          \\
kfp & Li$_2$BaP$_2$O$_7$        \\
\hline
\end{tabular}
\end{center}
\end{table}

\subsection{AIMD Parameters}
AIMD simulations were performed using the Vienna \textit{Ab initio} Simulation Package (VASP) \citep{kresse1996computational, kresse1996efficient, kresse1993ab, kresse1999ultrasoft}, employing a minimal $\Gamma$-centered $1\times1\times1$ $k$-point grid to reduce computational cost and an energy cutoff of 520~eV. All calculations were non-spin-polarized and used the Perdew--Burke--Ernzerhof (PBE) exchange--correlation functional. Projector Augmented Wave (PAW) pseudopotentials were selected following the Materials Project recommended mapping, as implemented in the \texttt{MPRelaxSet} VASP input set from \texttt{pymatgen} \citep{ong2013python, jain2011high}. These choices were made to maximize consistency with the DFT data on which the Mace \texttt{medium-mpa-0} model was trained, which was computed at the PBE+U level with PAW pseudopotentials. An exception that we note is that no Hubbard U correction was applied here, which in practice only affects \ce{LiVO3} among the materials studied. Another notable difference is that portions of both training datasets include spin-polarized calculations, whereas all AIMD runs reported here are non-spin-polarized. Finally, the training data was generated using dense k-point meshes, while the present calculations use only the $\Gamma$ point, which is well-justified for the large supercells employed but may introduce small systematic offsets in absolute energies and forces relative to the training references.
Regarding the supercell size, exceptions to minimum dimensions of $10 \mathrm{\AA} \times 10 \mathrm{\AA} \times 10 \mathrm{\AA}$ were made for AIMD simulations of materials \ce{Li5GaO4}, \ce{LiTi2(PO4)3}, and \ce{LiZr2(PO4)3}, for which only primitive cells were used to keep the number of atoms computationally feasible, and for \ce{Li6CuB4O10}, where one lattice dimension was not expanded beyond 10~\AA. In all cases, the remaining dimensions were at least 8~\AA. With 64 CPUs available per node, the number of nodes was determined as $n_{\rm nodes} = \max\big(1, n_{\rm atoms} // 64 )$, which in practice always resulted in a single node.

\label{app:aimd+ft}

\subsection{Additional analysis}
\label{app:mace_vs_dft+err}
\begin{figure}[h]
\centering
\includegraphics[width=0.5\linewidth]{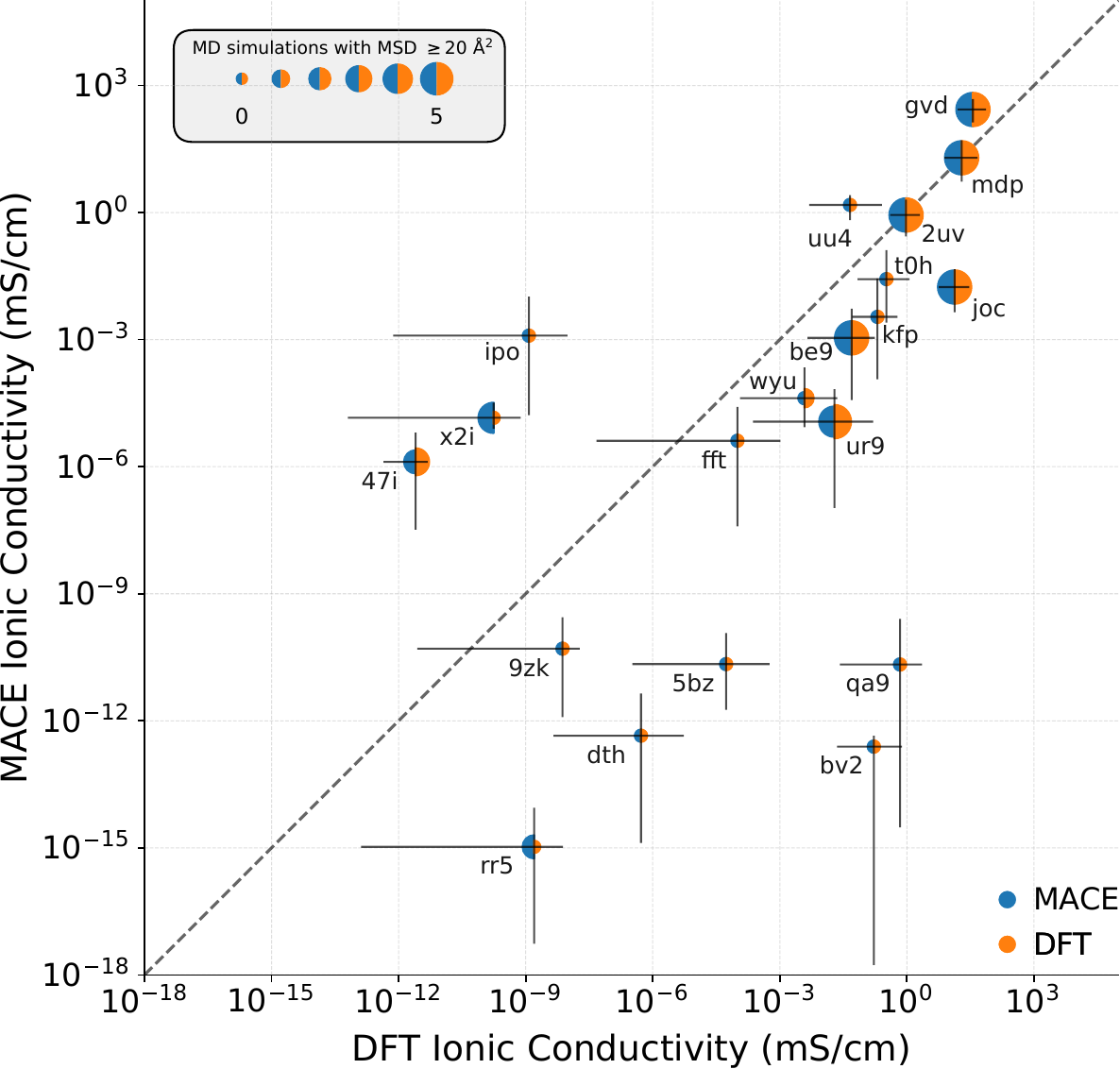}
\caption{Parity plot comparing ionic conductivity computed using DFT and MACE foundation model. The size of the half circles reflect the number of simulations (at 5 different temperatures) that reached at least $\mathrm{MSD} = 20~\mathrm{\AA}^2$.} 
\label{fig:mace_vs_dft}
\end{figure}

In Figure~\ref{fig:mace_vs_dft}, we present a parity plot comparing DFT and MACE predictions. In principle, DFT can be regarded as a reference for MACE given that the latter is trained on DFT calculations. However, it should be noted that MACE is trained on DFT data generated using computational settings that differ from those employed in this work. 

\begin{figure}[h]
\centering
\includegraphics[width=0.98\linewidth]{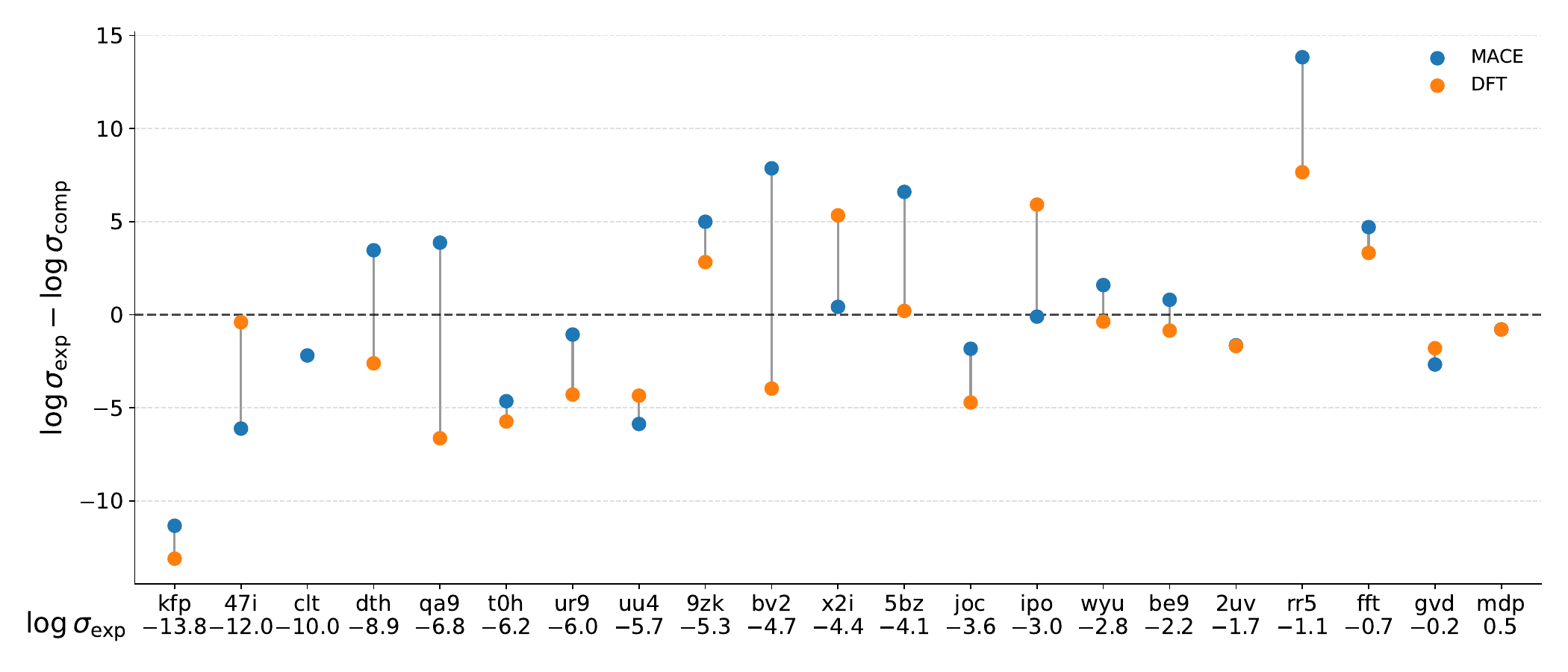}
\caption{
Logarithmic error between experimental and computed ionic conductivities for each material. The horizontal dashed line indicates perfect agreement. The x-axis lists materials equally spaced for clarity, with experimental $\log_{10}\sigma_{\rm exp} \,[\mathrm{mS/cm}]$ values shown below the labels. Materials are ordered by increasing experimental conductivity.
}
\label{fig:err_mace_dft_exp}
\end{figure}

Figure~\ref{fig:err_mace_dft_exp} shows the prediction errors of MACE and DFT with respect to experimental measurements.

\end{document}